\documentclass[natbib]{sigplanconf}
\usepackage{amsmath}
\usepackage[pdftex]{graphicx,color}
\usepackage{stmaryrd}
\usepackage{amssymb}

\usepackage{natbib}
\bibpunct();A{},
\let\cite=\citep

\newtheorem{theorem}{Theorem}

\newtheorem{lemma}{Lemma}
\newtheorem{proposition}{Proposition}

\newcommand\True{\ensuremath{\mbox{\tt True}}}
\newcommand\False{\ensuremath{\mbox{\tt False}}}
\newcommand\TT{\ensuremath{\mbox{\tt tt}}}
\newcommand\FF{\ensuremath{\mbox{\tt ff}}}
\newcommand\Not{\ensuremath{\mbox{\tt Not}}}
\newcommand\Copy{\ensuremath{\mbox{\tt Copy}}}
\newcommand\Implies{\ensuremath{\mbox{\tt Implies}}}

\newcommand\AND{\ensuremath{\mbox{\tt And}}}
\newcommand\Or{\ensuremath{\mbox{\tt Or}}}
\newcommand\Widget{\ensuremath{\mbox{\tt Widget}}}
\newcommand\Extract{\ensuremath{\mbox{\tt Extract}}}
\newcommand\Null{\ensuremath{\mbox{\tt Null}}}

\newcommand\Zero{\ensuremath{\mathbf 0}}  
\newcommand\One{\ensuremath{\mathbf 1}}

\newcommand\Lab{\mathbf{Lab}}

\newcommand\CEnv{\mathbf{CEnv}}
\newcommand\Cache{\mathbf{Cache}}
\newcommand\Var{\mathbf{Var}}
\newcommand\Val{\mathbf{Val}}
\newcommand\Exp{\mathbf{Exp}}
\newcommand\Term{\mathbf{Term}}
\newcommand\cache{\widehat{\mathsf{C}}}
\newcommand\ecache{\mathsf{C}}          

\newcommand\ce{ce}
\newcommand\restrict{\ensuremath{\!\upharpoonright\!}}
\newcommand\kmodels[2]{\models^{#1}_{#2}}
\newcommand\fv[1]{\ensuremath{\mathbf{fv}(#1)}}
\newcommand\ev[3]{\ensuremath{\mathcal{E}\sem{#1}^{#2}_{#3}}}
\newcommand\av[3]{\ensuremath{\mathcal{A}\sem{#1}^{#2}_{#3}}}
\newcommand\sem[1]{\ensuremath{\llbracket #1\rrbracket}}
\newcommand\exptime{{\sc exptime}}
\newcommand\ptime{{\sc ptime}}
\newcommand\np{{\sc np}}

\begin{document}

\conferenceinfo{ICFP'08,} {September 22--24, 2008, Victoria, BC, Canada.}
\CopyrightYear{2008}
\copyrightdata{978-1-59593-919-7/08/09}

\title{Deciding $k$CFA is complete for EXPTIME\thanks{Work supported under NSF Grant CCF-0811297.}}
\authorinfo{David {Van Horn} \and Harry G.~Mairson}
	   {Brandeis University}
           {\{dvanhorn,mairson\}@cs.brandeis.edu}
\maketitle

\begin{abstract}
We give an exact characterization of the computational complexity of
the $k$CFA hierarchy.  For any $k > 0$, we prove that the control flow decision
problem is complete for deterministic exponential time.  This theorem validates
empirical observations that such control flow analysis is intractable.  It also provides
more general insight into the complexity of abstract interpretation.
\end{abstract}




\section{Introduction}

Flow analysis \cite{jones-81,sestoft-88,shivers-88,midtgaard-07} is
concerned with the sound approximation of run-time values at compile
time.  This analysis gives rise to natural decision problems such as:
{\em does expression $e$ possibly evaluate to value $v$ at run-time?}
or {\em does function $f$ possibly get applied at call site
$\ell$?}\footnote{There is a bit more nuance to the question, which
will be developed later.}  The most approximate analysis always
answers {\em yes}.  This crude ``analysis'' takes no resources to
compute, and is useless.  In complete contrast, the most precise
analysis only answers {\em yes} by running the program to find that
out.  While such information is surely useful, the cost of this
analysis is likely prohibitive, requiring intractable or unbounded
resources.  Practical flow analyses occupy a niche between these
extremes, and their {\em expressiveness} can be characterized by the
computational resources required to compute their results.

Examples of simple yet useful flow analyses include Shivers' 0CFA
\citeyearpar{shivers-88} and Henglein's simple closure analysis
\citeyearpar{henglein92d}, which are {\em monovariant}---functions
that are closed over the same $\lambda$-expression are identified.
Their expressiveness is characterized by the class \ptime\
\cite{vanhorn-mairson-icfp07, vanhorn-mairson-sas08}.  More precise
analyses can be obtained by incorporating context-sensitivity to
distinguish multiple closures over the same $\lambda$-term.  The
$k$CFA hierarchy uses the last $k$ calling contexts to distinguish
closures, resulting in ``finer grained approximations, expending more
work to gain more information'' \cite{shivers-88,shivers-phd}.

The increased precision comes with an empirically observed increase in
cost. As Shivers noted in his retrospective on the $k$CFA work
\citeyearpar{shivers-sigplan04}:
\begin{quote}\it
It did not take long to discover that the basic analysis, for any $k >
0$, was intractably slow for large programs. In the ensuing years,
researchers have expended a great deal of effort deriving clever ways
to tame the cost of the analysis.
\end{quote}

A fairly straightforward calculation---see, for example,
\citet{nielson-nielson-hankin}---shows that 0CFA can be computed in
polynomial time, and for any $k>0$, $k$CFA can be computed in
exponential time.  These naive upper bounds suggest that the $k$CFA
hierarchy is essentially {\em flat}; researchers subsequently
``expended a great deal of effort'' trying to improve
them.\footnote{Even so, there is a big difference between algorithms
that run in $2^n$ and $2^{n^2}$ steps, though both are nominally in
\exptime.}  For example, it seemed plausible (at least, to us) that
the $k$CFA problem
could be in \np\ by {\em guessing} flows appropriately during analysis.

In this paper, we show that the naive algorithm is essentially the
best one, and that the {\em lower} bounds are what needed improving.
We prove that for all $k>0$, computing the $k$CFA analysis requires
(and is thus complete for) deterministic exponential time.  There is, in the
worst case---and plausibly, in practice---no way to tame the cost of the analysis.
Exponential time is required.

Who cares, and why should this result matter to functional programmers?

This result concerns a fundamental and ubiquitous static analysis {\em
of} functional programs.  The theorem gives an analytic, scientific
characterization of the expressive power of $k$CFA.  As a consequence,
the {\em empirically observed} intractability of the cost of this
analysis can be understood as being {\em inherent in the approximation
problem being solved}, rather than reflecting unfortunate gaps in our
programming abilities.

Good science depends on having relevant theoretical understandings of
what we observe empirically in practice---otherwise, we devolve to an
unfortunate situation resembling an old joke about the difference
between geology (the observation of physical phenomena that cannot be
explained) and geophysics (the explanation of physical phenomena that
cannot be observed).  Or worse---scientific aberrations such as
alchemy, or Ptolemaic epicycles in astronomy.

This connection between theory and experience contrasts with the
similar result for ML type inference \cite{mairson-popl90}: we confess
that while the problem of recognizing ML-typable terms is complete for
exponential time, programmers have happily gone on programming.  It is
likely that their need of higher-order procedures, essential for the
lower bound, is not considerable.  But static flow analysis really has
been costly, and our theorem explains why.

The theorem is proved {\em by} functional programming.
We take the view that the analysis itself is a functional programming
language, albeit with implicit bounds on the available computational
resources.  Our result harnesses the approximation inherent in $k$CFA
as a computational tool to hack exponential time Turing machines
within this unconventional language.  The hack used here is completely unlike
the one used for the ML analysis, which depended on complete developments
of {\tt let}-redexes.  The theorem we prove in this paper uses approximation
in a way that has little to do with normalization.

\section{Preliminaries}
\subsection{Instrumented interpretation}
\label{sec:instrumented}

$k$CFA can be thought of as an abstraction (in the sense of a
computable approximation) to an instrumented interpreter, which not
only evaluates a program, but records a history of {\em flows}.  Every
time a subterm evaluates to a value, every time a variable is bound to
a value, the {\em flow} is recorded.  Consider a simple example, where
$e$ is closed and in normal form:
\begin{displaymath}
(\lambda x. x) e
\end{displaymath}
We label the term to index its constituents:
\begin{displaymath}
((\lambda x. x^0)^1 e^2)^3
\end{displaymath}

The interpreter will first record all the flows for evaluating $e$
(there are none, since it is in normal form), then the flow of $e$'s
value (which is $e$, closed over the empty environment) into label $2$
($e$'s label) is recorded.

This value is then recorded as flowing into the binding of $x$.  The
body of the $\lambda x$ expression is evaluated under an extended
environment with $x$ bound to the result of evaluating $e$.  Since it
is a variable occurrence, the value bound to $x$ is recorded as
flowing into this variable occurrence, labeled $0$.  Since this is
the result of evaluating the body of the function, it is recorded as
flowing out of the $\lambda$-term labeled $1$.  And again, since this
is the result of the application, the result is recorded for label
$3$.

The flow history is recorded in a {\em cache}, $\ecache$, which maps
labels and variables to values.  If the cache maps a variable to a
value, $\ecache(x) = v$, it means that during evaluation of the
program, the variable $x$ was bound to the value $v$ at some point.  If
the cache maps a label to a value, $\ecache(\ell) = v$, it means that the
subexpression with that label evaluated to that value.

Of course, a variable may be bound to any number of values during the
course of evaluation.  Likewise, a subexpression that occurs once
syntactically may evaluate to any number of values during evaluation.
So asking about the flows of a subexpression is ambiguous without
further information.  Our simple example does not reflect this possible ambiguity, but
consider the following example, where \True\ and \False\ are closed and
in normal form:
\begin{displaymath}
(\lambda f.f(f\; \True)) (\lambda y. \False)
\end{displaymath}
During evaluation, $y$ gets bound to both \True\ and \False---asking
``what was $y$ bound to?'' is ambiguous.  But let us label the
applications in our term:
\begin{displaymath}
((\lambda f.(f(f\; \True)^1)^2) (\lambda y. \False))^3
\end{displaymath}
Notice that $y$ is bound to different values within different
contexts.  That is, $y$ is bound \True\ when evaluating the
application labeled 1, and to \False\ when evaluating the application
labeled 2.  Both of these occur while evaluating the outermost
application, labeled 3.  A string of these application labels, called
a {\em contour}, uniquely describes the {\em context} under which a
subexpression evaluates.
\begin{eqnarray*}
3\cdot 2\cdot 1 & \mbox{describes} & ((\lambda f.(f [\;]^1)^2) (\lambda y.\False))^3\\
3\cdot 2 & \mbox{describes} & ((\lambda f.[\;]^2) (\lambda y.\False))^3
\end{eqnarray*}

So a question about what a subexpression evaluates to {\em within a
given context} has an unambiguous answer.  The interpreter, therefore,
maintains an environment that maps each variable to a description of
the context in which it was bound.  Similarly, flow questions about a
particular subexpression or variable binding must be accompanied by a
description of a context.  Returning to our example, we would have
$\ecache(y,3\cdot 2\cdot 1) = \True$ and $\ecache(y,3\cdot 2) =
\False$.

Typically, values are denoted by {\em closures}---a $\lambda$-term
together with an environment mapping all free variables in the term to
values.\footnote{We abused syntax above by writing the closure of a
  closed term as the term itself, rather than
  $\langle\True,\emptyset\rangle$, for example.  We continue with the
  abuse.}  But in the instrumented interpreter, rather than mapping
variables to values, environments map a variable to a {\em
  contour}---the sequence of labels which describes the context of
successive function applications in which this variable was
bound:\footnote{All of the syntactic categories are implicitly
  understood to be restricted to the finite set of terms, labels,
  variables, etc.~that occur in the {\em program of interest}---the
  program being analyzed.  As a convention, programs are assumed to
  have distinct bound variable names and labels.}
\begin{displaymath}
\begin{array}{cclcll}
\delta & \in & \Delta & = &  \Lab^{n}  & \mbox{contour}\\
\ce    & \in & \CEnv  & = & \Var \rightarrow \Delta & \mbox{contour environment}
\end{array}
\end{displaymath}
By consulting the cache, we can then retrieve the value.  So if under
typical evaluation, a term labeled $\ell$ evaluates to $\langle
\lambda x.e,\rho\rangle$ within a context described by a string of
application labels, $\delta$, then we will have $\ecache(\ell,\delta)
= \langle \lambda x . e, \ce\rangle$, where the contour environment
$ce$, like $\rho$, closes $\lambda x.e$.  But unlike $\rho$, it maps
each variable to a contour describing the context in which the
variable was bound.  So if $\rho(y) = \langle \lambda
z.e',\rho'\rangle$, then $\ecache(y,\ce(y)) = \langle \lambda
z.e',\ce'\rangle$, where $\rho'$ is similarly related to $\ce'$.

We can now write the instrumented evaluator.  The syntax of the
language is given by the following grammar:
\begin{displaymath}
\begin{array}{l@{\quad}l@{\quad}l}
\Exp & e ::= t^\ell & \mbox{expressions (or labeled terms)}\\
\Term & t ::= x\ |\ e\;e\ |\ \lambda x.e & \mbox{terms (or unlabeled expressions)}
\end{array}
\end{displaymath}

$\ev{t^\ell}{\ce}{\delta}$ evaluates $t$ and writes the result into
the table $\ecache$ at location $(\ell,\delta)$.  The notation
$\ecache(\ell,\delta) \leftarrow v$ means that the cache is updated so that
$\ecache(\ell,\delta) = v$.  The notation $\delta\ell$ denotes the
concatenation of contour $\delta$ and label $\ell$.
\begin{displaymath}
\begin{array}{lcl}
\ev{x^\ell}{\ce}{\delta} & = & \ecache(\ell,\delta) \leftarrow \ecache(x,\ce(x))\\
\ev{(\lambda x.e)^\ell}{\ce}{\delta} & = & \ecache(\ell,\delta) \leftarrow \langle\lambda x.e,\ce'\rangle\\
\ &\ & \quad\mbox{where } \ce' = \ce \restrict \fv{\lambda x.e}\\
\ev{(t^{\ell_1} t^{\ell_2})^\ell}{\ce}{\delta} & = & \ev{t^{\ell_1}}{\ce}{\delta}; \ev{t^{\ell_2}}{\ce}{\delta};\\
\ &\ & \mbox{let }\langle\lambda x.t^{\ell_0},\ce'\rangle = \ecache(\ell_1,\delta)\mbox{ in}\\
\ &\ & \quad\ecache(x,\delta\ell) \leftarrow \ecache(\ell_2,\delta);\\
\ &\ & \quad\ev{t^{\ell_0}}{\ce'[x\mapsto\delta\ell]}{\delta\ell}\\
\ &\ & \quad\ecache(\ell,\delta)\leftarrow\ecache(\ell_0,\delta\ell)
\end{array}
\end{displaymath}
The cache constructed by
\begin{displaymath}
\ev{((\lambda f.(f(f\; \True)^1)^2) (\lambda y. \False))^3}{\emptyset}{\epsilon}
\end{displaymath}
includes the following entries:
\begin{eqnarray*}
\ecache(f,3) & = & \lambda y.\False\\
\ecache(y,3\cdot 2\cdot 1) & = & \True\\
\ecache(1,3\cdot 2) & = & \lambda y.\False\\
\ecache(y,3\cdot 2) & = & \False
\end{eqnarray*}

In a more declarative style, we can write a specification of
{\em acceptable caches}---a cache is acceptable iff it records all of the
flows which occur during evaluation.  The smallest cache satisfying
this acceptability relation is the one that is computed by the above interpreter.
\begin{displaymath}
\begin{array}{lcl}
\ecache \kmodels{\ce}{\delta} x^\ell & 
\mbox{iff} & 
\ecache(x,\ce(x)) = \ecache(\ell,\delta) \\
\ecache \kmodels{\ce}{\delta}  (\lambda x .e)^\ell & 
\mbox{iff} & 
\langle\lambda x .e,\ce'\rangle = \ecache(\ell,\delta) \\
\ & \ & \quad  \mbox{where } \ce' = ce \restrict \fv{\lambda x.e}\\
\ecache \kmodels{\ce}{\delta} (t^{\ell_1}\ t^{\ell_2})^\ell & 
\mbox{iff} & 
\ecache\kmodels{\ce}{\delta} t^{\ell_1} \wedge
\ecache\kmodels{\ce}{\delta} t^{\ell_2} \;\wedge \\
\ & \ & \mbox{let }\langle\lambda x.t^{\ell_0},\ce'\rangle = \ecache(\ell_1,\delta) \mbox{ in}\\
\ & \ & \quad\ecache(\ell_2,\delta) = \ecache(x,\delta\ell) \;\wedge \\
\ & \ & \quad\ecache\kmodels{\ce'[x\mapsto\delta\ell]}{\delta\ell} t^{\ell_0}\; \wedge  \\
\ & \ & \quad\ecache(\ell_0,\delta\ell) = \ecache(\ell,\delta) 
\end{array}
\end{displaymath}

Clearly, because constructing a cache $\ecache$ is equivalent to
evaluating a program, such a cache is not effectively computable.
The next section describes $k$CFA as a computable {\em approximation.}

\subsection{An abstract interpreter}

$k$CFA is a computable approximation to this instrumented interpreter.
Rather than constructing an {\em exact} cache $\ecache$, it constructs
an {\em abstract} cache $\cache$, which maps labels and variables, not
to values, but to {\em sets} of {\em abstract values}.  
\begin{displaymath}
\begin{array}{cclcll}
\hat{v}      & \in & \widehat{\Val}   & = & \mathcal{P}(\Term \times \CEnv) \\
\cache & \in & \widehat{\Cache} & = & (\Lab+\Var) \times \Delta \rightarrow \widehat{\Val} \\
\end{array}
\end{displaymath}

Approximation arises from contours being bounded at length $k$.  If
during the course of instrumented evaluation, the length of the
contour would exceed length $k$, then the $k$CFA abstract interpreter
will truncate it to length $k$.  In other words, only a partial
description of the context can be given, which results in ambiguity.
A subexpression may evaluate to two distinct values, but within
contexts which are only distinguished by $k+1$ labels.  Questions
about which value the subexpression evaluates to can only supply $k$
labels, so the answer must be {\em both}, according to a sound
approximation.

When applying a function, there is now a set of possible closures
that flow into the operator position.  Likewise, there can be a
multiplicity of arguments.  What is the interpreter to do?  The abstract 
interpreter applies all possible closures to all possible arguments.

The abstract interpreter, the imprecise analog of $\mathcal{E}$, is then:
\begin{displaymath}
\begin{array}{lcl}
\av{x^\ell}{\ce}{\delta} & = & \cache(\ell,\delta) \leftarrow \cache(x,\ce(x))\\
\av{(\lambda x.e)^\ell}{\ce}{\delta} & = & \cache(\ell,\delta) \leftarrow \{\langle\lambda x.e,\ce'\rangle\}\\
\ &\ & \quad\mbox{where } \ce' = \ce \restrict \fv{\lambda x.e}\\
\av{(t^{\ell_1} t^{\ell_2})^\ell}{\ce}{\delta} & = & \av{t^{\ell_1}}{\ce}{\delta}; \av{t^{\ell_2}}{\ce}{\delta};\\
\ &\ &\mbox{foreach }\langle\lambda x.t^{\ell_0},\ce'\rangle\in\cache(\ell_1,\delta):\\
\ &\ & \quad\cache(x,\lceil\delta\ell\rceil_k) \leftarrow \cache(\ell_2,\delta);\\
\ &\ & \quad\av{t^{\ell_0}}{\ce'[x\mapsto\lceil\delta\ell\rceil_k]}{\lceil\delta\ell\rceil_k};\\
\ &\ & \quad\cache(\ell,\delta)\leftarrow\cache(\ell_0,\lceil\delta\ell\rceil_k)
\end{array}
\end{displaymath}
We write $\cache(\ell,\delta) \leftarrow \hat{v}$ to indicate an updated cache
where $(\ell,\delta)$ maps to 
$\cache(\ell,\delta) \cup \hat{v}$.  The notation
$\lceil\delta\rceil_k$ denotes $\delta$ truncated to the rightmost (i.e., most recent)
$k$ labels.  

Compared to the exact evaluator, contours similarly distinguish
evaluation within contexts described by as many as $k$ application
sites: beyond this, the distinction is blurred.  The imprecision of
the analysis requires that $\mathcal{A}$ be iterated until the cache
reaches a fixed point, but care must taken to avoid looping in an
iteration since a single iteration of $\av{e}{\ce}{\delta}$ may in
turn make a recursive call to $\av{e}{\ce}{\delta}$ under the same
contour and environment.  This care is the algorithmic analog of
appealing to the coinductive hypothesis in judging an analysis
acceptable.  These judgment rules are given below.

An acceptable $k$-level control flow analysis for an expression $e$ is
written $\cache\kmodels{\ce}{\delta} e$, which states that $\cache$ is
an acceptable analysis of $e$ in the context of the current
environment $\ce$ and current contour $\delta$ (for the top level
analysis of a program, these will both be empty).  

Just as we did in the previous section, we can write a specification
of acceptable caches rather than an algorithm that computes.  The
resulting specification is what is found, for example, in
\citet{nielson-nielson-hankin}:
\begin{displaymath}
\begin{array}{lcl}
\cache \kmodels{\ce}{\delta} x^\ell & \mbox{iff} & \cache(x,\ce(x)) \subseteq \cache(\ell,\delta)\\
\cache \kmodels{\ce}{\delta} (\lambda x .e)^\ell & \mbox{iff} & \langle\lambda x .e,\ce'\rangle \in \cache(\ell,\delta)\\
\ & \ & \quad  \mbox{where } \ce' = ce \restrict \fv{\lambda x.e}\\
\cache \kmodels{\ce}{\delta} (t^{\ell_1}\ t^{\ell_2})^\ell & \mbox{iff} & \cache\kmodels{\ce}{\delta} t^{\ell_1} \wedge
\cache\kmodels{\ce}{\delta} t^{\ell_2} \wedge \\
\ &\ & \forall \langle\lambda x.t^{\ell_0},\ce'\rangle \in \cache(\ell_1,\delta) : \\
\ &\ & \quad\cache(\ell_2,\delta) \subseteq \cache(x,\lceil\delta\ell\rceil_k) \wedge\\
\ &\ & \quad\cache\kmodels{ce'[x \mapsto \lceil\delta\ell\rceil_k]}{\lceil\delta\ell\rceil_k} t^{\ell_0} \wedge  \\
\ &\ & \quad\cache(\ell_0,\lceil\delta\ell\rceil_k) \subseteq \cache(\ell,\delta)
\end{array}
\end{displaymath}

The acceptability relation is given by the greatest fixed point of the
functional defined according to the above clauses---and we are
concerned only with least solutions.\footnote{To be precise, we take
as our starting point {\em uniform} $k$CFA
\cite{nielson-nielson-popl97} rather than a $k$CFA in which
$\widehat{\Cache} = (\Lab \times \CEnv) \rightarrow \widehat{\Val}$.
The differences are immaterial for our purposes.  See
\citet{nielson-nielson-hankin} for details and a discussion on the use
of coinduction in specifying static analyses.}

\subsection{Complexity of abstract interpretation}


What is the difficulty of computing within this hierarchy?  What are
the sources of approximation that render such analysis (in)tractable?
We consider these questions by analyzing the complexity of the following decision
problem:

\begin{description}
\item[Control Flow Problem:] Given an expression $e$, an abstract
value $\{v\}$, and a pair $(\ell,\delta)$, is $v\in\cache(\ell,\delta)$ in
the flow analysis of $e$?
\end{description}

Obviously, we are interested in the complexity of control flow
analysis, but our investigation also provides insight into a more
general subject: the complexity of computing via abstract
interpretation.  It stands to reason that as the computational domain
becomes more refined, so too should computational complexity.  In this
instance, the domain is the size of the abstract cache $\cache$ and
the values (namely, {\em closures}) that can be stored in the cache.
As the table size and number of closures increase\footnote{Observe
  that since closure environments map free variables to contours, the
  number of closures increases when we increase the contour length
  $k$.}, so too should the complexity of computation.  From a theoretical
perspective, we would like to understand better the tradeoffs between these
various parameters.

\section{Linearity and Boolean logic}

It is straightforward to observe that in a {\em linear}
$\lambda$-term, where each variable occurs at most once, each
abstraction $\lambda x.e$ can be applied to at most one argument, and
hence the abstracted value can be bound to at most one argument.
(Note that this observation is clearly untrue for the {\em nonlinear}
$\lambda$-term $(\lambda f.f (a (f b))) (\lambda x.x)$, as $x$ is
bound to $b$, and also to $ab$.)  Generalizing this observation, analysis of
a linear $\lambda$-term coincides exactly with its evaluation:

\begin{lemma}
For any closed, linear expression $e$, 
$\ev{e}{\emptyset}{\epsilon} = \av{e}{\emptyset}{\epsilon}$, and thus
$\ecache=\cache$.
\end{lemma}
A detailed proof of this lemma appears in
\citet{vanhorn-mairson-sas08}.

A natural and expressive class of such
linear terms are the ones which implement Boolean logic.  When we
analyze the coding of a Boolean circuit and inputs to it, the Boolean
output will flow to a predetermined place in the (abstract) cache.  By
placing that value in an appropriate context, we construct an instance of
the control flow problem: a function $f$ flows to a call site $a$ iff the
Boolean output is $\True$.  

Since we have therefore reduced the circuit value problem
\cite{ladner-75}, which is complete for \ptime, to an instance of the 0CFA control flow problem,
we conclude that the control flow problem is \ptime-hard.  Further, as
0CFA can be computed in polynomial time, the control flow problem for
0CFA is \ptime-complete.

We use some standard syntactic sugar for constructing and
deconstructing pairs:
\begin{eqnarray*}
\langle u,v\rangle & \equiv & \lambda z . zuv\\
\mbox{let }\langle x, y\rangle = p\mbox{ in }e & \equiv &
p (\lambda x.\lambda y. e)
\end{eqnarray*}
Booleans are built out of constants \TT\ and \FF, which are the
identity and pair swap function, respectively:
\begin{displaymath}
\begin{array}{rcl@{\qquad}rcl}
\TT & \equiv & \lambda p . \mbox{let }\langle x,y \rangle = p\mbox{ in } \langle x, y\rangle &
\True & \equiv &\langle \TT,\FF \rangle\\
\FF & \equiv & \lambda p . \mbox{let }\langle x,y \rangle = p\mbox{ in } \langle y, x\rangle &
\False & \equiv &\langle \FF,\TT \rangle
\end{array}
\end{displaymath}

The simplest connective is \Not, which is an inversion on pairs, like
\FF.  A {\em linear} copy connective is defined as:
\begin{eqnarray*}
\Copy & \equiv & \lambda b.\mbox{let }\langle u,v\rangle = b\mbox{ in }
\langle u\langle \TT,\FF\rangle, v\langle \FF,\TT\rangle\rangle
\end{eqnarray*}
The coding is easily explained: suppose $b$ is \True, then $u$ is identity and $v$
twists; so we get the pair $\langle\True,\True\rangle$.  Suppose $b$
is \False, then $u$ twists and $v$ is identity; we get
$\langle\False,\False\rangle$.  We write $\Copy_n$ to mean $n$-ary
fan-out---a straightforward extension of the above.

Now we define truth-table implication:
\begin{displaymath}
\begin{array}{rcl}
\Implies & \equiv & \lambda b_1.\lambda b_2.\\
\ & \ & \quad\mbox{let }\langle u_1,v_1\rangle = b_1\mbox{ in}\\
\ & \ & \quad\mbox{let }\langle u_2,v_2\rangle = b_2\mbox{ in}\\
\ & \ & \quad\mbox{let }\langle p_1,p_2\rangle = u_1\langle u_2, \TT\rangle \mbox{ in}\\
\ & \ & \quad\mbox{let }\langle q_1,q_2\rangle = v_1\langle \FF, v_2\rangle \mbox{ in}\\
\ & \ & \qquad\langle p_1, q_1 \circ p_2 \circ q_2 \circ \FF \rangle
\end{array}
\end{displaymath}
Notice that if $b_1$ is \True, then $u_1$ is \TT, so $p_1$ is \TT\ iff
$b_2$ is \True.  And if $b_1$ is \True, then $v_1$ is \FF, so $q_1$ is
\FF\ iff $b_2$ is \False.  On the other hand, if $b_1$ is \False,
$u_1$ is \FF, so $p_1$ is \TT, and $v_1$ is \TT, so $q_1$ is \FF.
Therefore $\langle p_1,q_1\rangle$ is \True\ iff $b_1 \supset b_2$,
and \False\ otherwise.\footnote{Or, if you prefer, $u_1 \langle
u_2,\TT\rangle$ can be read as ``if $u_1$, then $u_2$ else
$\TT$''---the if-then-else description of the implication $u_1\supset
u_2$ ---and $v_1 \langle \FF, v_2\rangle$ as its deMorgan dual
$\neg(v_2\supset v_1)$.  Thus $\langle p_1,q_1\rangle$ is the answer
we want---and we need only dispense with the ``garbage'' $p_2$ and
$q_2$.  DeMorgan duality ensures that one is $\TT$, and the other is
$\FF$ (though we do not know which), so they always compose to $\FF$.}

However, simply returning $\langle p_1,q_1\rangle$ violates linearity since
$p_2,q_2$ go unused.  We know that $p_2 = \TT$ iff $q_2 = \FF$ and
$p_2 = \FF$ iff $q_2 = \TT$.  We do not know which is which, but
clearly $p_2 \circ q_2 = \FF\circ\TT = \TT\circ\FF = \FF$.  Composing $p_2\circ
q_2$ with $\FF$, we are guaranteed to get $\TT$.  Therefore $q_1
\circ p_2 \circ q_2 \circ \FF = q_1$, and we have used all bound
variables exactly once.  The \AND\ and \Or\ connectives are defined
similarly (as in \citet{vanhorn-mairson-icfp07}).

\subsection{The Widget}

Consider a Boolean circuit coded as a program: it can only evaluate to
a (coded) true or false value, but a flow analysis identifies terms by
label, so it is possible several different \True\ and \False\ terms
flow out of the program.  But our decision problem is defined with
respect to a particular term.  What we want is to use flow analysis to
answer questions like ``does this program (possibly) evaluate to a
true value?''  We use The Widget to this effect.  It is a term
expecting a boolean value.  It evaluates as though it were the
identity function on Booleans, $\Widget\; b = b$, but it induces a
specific flow we can ask about. If a true value flows out of $b$, then
$\True_W$ flows out of $\Widget\; b$.  If a false value flows out of
$b$, then $\False_W$ flows out of $\Widget\; b$, where $\True_W$ and
$\False_W$ are distinguished terms, and the only possible terms that
can flow out.  We usually drop the subscripts and say ``does \True\
flow out of $\Widget\; b$?''  without much ado.\footnote{This \Widget\
is affine, but this is only for simplicity in presentation.  A
non-affine widget is given \citet{vanhorn-mairson-icfp07}.}
\begin{eqnarray*}
\Widget & \equiv & \lambda b.\mbox{let }\langle u,v\rangle=b\mbox{ in }
\pi_1 (u\langle\True_W,\False_W\rangle)
\end{eqnarray*}

Because the circuit value problem is complete for \ptime, we
conclude \cite{vanhorn-mairson-icfp07}:

\begin{theorem}
Deciding the control flow problem for 0CFA is complete for \ptime.
\end{theorem}

\section{\ Nonlinearity and Cartesian products:\\ \qquad a toy calculation, with insights}

A good proof has, at its heart, a small and simple idea that makes it
work.  For our proof, the key idea is how the approximation of
analysis can be {\em leveraged} to provide computing power {\em above
and beyond} that provided by evaluation.  The difference between the
two can be illustrated by the following term:
\begin{displaymath}
\begin{array}{l}
(\lambda f. (f\;\mbox{\tt True}) (f\;\mbox{\tt False}))\\
(\lambda x.\Implies\, x\, x)
\end{array}
\end{displaymath}
Consider evaluation: Here $\Implies\, x\, x$ (a tautology) is evaluated
twice, once with $x$ bound to \True, once with $x$ bound to \False.
But in both cases, the result is \True.  Since $x$ is bound to \True\
or \False\, both occurrences of $x$ are bound to \True\ or to
\False---but it is never the case, for example, that the first
occurrence is bound to \True, while the second is bound to \False.
The values of each occurrence of $x$ is dependent on the other.

On the other hand, consider what flows out of $\Implies\,x\, x$ according
1CFA: both \True\ and \False.  Why? The approximation incurs analysis
of $\Implies\,x\, x$ for $x$ bound to \True\ and \False, but it considers
{\em each occurrence of $x$ as ranging over \True\ and \False,
independently}.  In other words, for the set of values bound to $x$,
we consider their {\em cross product} when $x$ appears non-linearly.
The approximation permits one occurrence of $x$ be bound to \True\
while the other occurrence is bound to \False; and somewhat alarmingly,
$\Implies\,\True\,\False$ causes \False\ to flow out.  Unlike in normal
evaluation, where within a given scope we know that multiple
occurrences of the same variable refer to the same value, in the
approximation of analysis, multiple occurrences of the same variable
range over {\em all} values that they are possible bound to {\em
independent of each other}.  

Now consider what happens when the program is expanded as follows:
\begin{displaymath}
\begin{array}{l}
(\lambda f. (f\,\mbox{\tt True}) (f\,\mbox{\tt False}))\\
(\lambda x. (\lambda p. p (\lambda u. p (\lambda v. \Implies\, u\, v))) (\lambda w.wx))
\end{array}
\end{displaymath}
Here, rather than pass $x$ directly to \Implies, we construct a unary
tuple $\lambda w.wx$.  The tuple is used non-linearly, so $p$ will
range over {\em closures} of $\lambda w.wx$ with $x$ bound to \True\
and \False, again, independently.

A closure can be approximated by an exponential number of values.  For
example, $\lambda w.wz_1z_2\dots z_n$ has $n$ free variables, so there
are an exponential number of possible environments mapping these
variables to program points (contours of length 1).  If we could apply
a Boolean function to this tuple, we would effectively be evaluating
all rows of a truth table; following this intuition leads to
\np-hardness of the 1CFA control flow problem.

Generalizing from unary to $n$-ary tuples in the above example, an
exponential number of closures can flow out of the tuple.  For a
function taking two $n$-tuples, we can compute the function on the
cross product of the exponential number of closures.

This insight is the key computational ingredient in simulating exponential
time, as we describe in the following section.

\section{The complexity of $k$CFA}
\subsection{Approximation and EXPTIME}

Recall the formal definition of a Turing machine: a 7-tuple
\[
\langle Q,\Sigma,\Gamma,\delta,q_0,q_a,q_r\rangle
\]
where $Q$, $\Sigma$, and $\Gamma$ are finite sets,  $Q$ is the set of
machine states (and $\{q_0,q_a,q_r\}\subseteq Q$),  $\Sigma$ is the
input alphabet, and $\Gamma$ the tape alphabet, where
$\Sigma\subseteq\Gamma$.  The states $q_0$, $q_a$, and $q_r$ are the
machine's initial, accept, and reject states, respectively.
The complexity class \exptime\ denotes the languages
that can be decided by a Turing machine in time  exponential in the
input length.  

Suppose we have a deterministic Turing machine $M$ that accepts or
rejects its input $x$ in time $2^{p(n)}$, where $p$ is a polynomial
and $n=|x|$.  We want to simulate the computation of $M$ on $x$ by
$k$CFA analysis of a $\lambda$-term $E$ dependent on $M,x,p$, where a
particular closure will flow to a specific program point iff $M$
accepts $x$.  It turns out that $k=1$ suffices to carry out this
simulation.  The construction, computed in logarithmic space, is
similar for all constant $k>1$ modulo a certain amount of padding.

\subsection{Coding machine IDs}

The first task is to code machine IDs.  Observe that each
value stored in the abstract cache $\cache$ is a {\em closure}---a
$\lambda$-abstraction, together with an environment for its free variables.
The number of such abstractions is bounded by the program size, as is the {\em domain}
of the environment---while the number of such {\em environments} is exponential in the program size.
(Just consider a program of size $n$ with, say, $n/2$ free variables mapped to only 2 program points
denoting bindings.)

Since a closure only has polynomial size, and a Turing machine ID has
exponential size, we represent the latter by splitting its information
into an exponential number of closures.  Each closure represents a
tuple $\langle T,S,H,C,b\rangle$, which can be read as
\begin{quote}\it
``At time $T$, Turing machine $M$ was in state $S$, the tape position
was at cell $H$, and cell $C$ held contents $b$.''
\end{quote}
$T$, $S$, $H$, and $C$ are blocks of bits ($\Zero \equiv \True$,
$\One\equiv\False$) of size polynomial in the input to the Turing machine.  As such, each block
can represent an exponential number of values.
A single machine ID is represented
by an exponential number of tuples (varying $C$ and $b$).  Each such tuple can in turn be
coded as a $\lambda$-term $\lambda w.wz_1z_2\cdots z_N$, where $N=O(p(n))$.

We still need to be able to generate an exponential number of closures for such an $N$-ary tuple.  The construction
is only a modest, iterative generalization of the construction in our toy calculation above:
\begin{displaymath}
\begin{array}{rcl}
\ & \ & (\lambda f_1.(f_1\;\Zero)(f_1\;\One))\\
\ & \ & (\lambda z_1.\\
\ & \ &\quad(\lambda f_2.(f_2\;\Zero)(f_2\;\One))\\
\ & \ &\quad(\lambda z_2.\\
\ & \ &\qquad\cdots\\
\ & \ &\quad\qquad(\lambda f_N. (f_N\;\Zero) (f_N\;\One))\\
\ & \ &\quad\qquad(\lambda z_N.((\lambda x.x)(\lambda w.wz_1z_2\cdots z_N))^{\ell})\cdots))
\end{array}
\end{displaymath}
In the final subterm $((\lambda x.x)(\lambda w.wz_1z_2\cdots
z_N))^{\ell}$, the function $\lambda x.x$ acts as a very important
form of {\em padding}.  Recall that this is $k$CFA with $k=1$---the
expression $(\lambda w.wz_1z_2\cdots z_N)$ is evaluated an exponential
number of times---to see why, normalize the term---but in each
instance, the contour is always $\ell$.  (For $k>1$, we would just
need more padding to evade the {\em polyvariance} of the flow
analyzer.)  As a consequence, each of the (exponential number of)
closures gets put in the {\em same} location of the abstract cache
$\cache$, while they are placed in unique, {\em different} locations
of the exact cache $\ecache$.  In other words, the approximation
mechanism of $k$CFA treats them as if they are all the same.  (That is
why they are put in the same cache location.)

\subsection{Transition function}

Now we define a binary transition function $\delta$, which does a {\em
piecemeal} transition of the machine ID.  The transition function is
represented by three rules, identified uniquely by the time stamps $T$
on the input tuples.

The first {\em transition rule} is used when the
tuples agree on the time stamp $T$, and the head and cell address of the
first tuple coincide:
\begin{displaymath}
\begin{array}{l}
\delta\langle T,S,H,H,b\rangle \langle T,S',H',C',b'\rangle\; = \\
\qquad\qquad\qquad 
\langle T+1, \delta_Q(S,b), \delta_{LR}(S,H,b), H, \delta_\Sigma(S,b) \rangle
\end{array}
\end{displaymath}
This rule {\em computes} the transition to the next ID.  The first
tuple has the head address and cell address coinciding, so it has all
the information needed to compute the next state, head movement, and
what to write in that tape cell.  The second tuple just marks that
this is an instance of the {\em computation} rule, simply indicated by
having the time stamps in the tuples to be identical.  The Boolean functions
$\delta_Q,\delta_{LR},\delta_\Sigma$ compute the next state, head position, and 
what to write on the tape.

The second {\em communication rule} is used when the tuples have time
stamps $T+1$ and $T$: in other words, the first tuple has information
about state and head position which needs to be communicated to every
tuple with time stamp $T$ holding tape cell information for an arbitrary such cell, as it
gets updated to time stamp $T+1$:
\begin{displaymath}
\begin{array}{l}
\delta\langle T+1,S,H,C,b\rangle \langle T,S',H',C',b'\rangle = \langle T+1, S, H, C',b'\rangle\\
\qquad\qquad\qquad\qquad\qquad\qquad\qquad\qquad\qquad 
(H'\not=C')
\end{array}
\end{displaymath}
(Note that when $H'=C'$, we have already written the salient tuple using the transition rule.)
This rule {\em communicates} state and head position (for the first
tuple computed with time stamp $T+1$, where the head and cell address
coincided) to all the other tuples coding the rest of the Turing
machine tape.  

Finally, we define a {\em catch-all rule}, mapping any other
pairs of tuples (say, with time stamps $T$ and $T+42$) to some
distinguished null value (say, the initial ID).  We need this rule just to make sure that
$\delta$ is a totally defined function.
\begin{displaymath}
\begin{array}{l}
\delta\langle T,S,H,C,b\rangle \langle T',S',H',C',b'\rangle\; =\ \Null\qquad\qquad\qquad\qquad\\
\qquad\qquad\qquad\qquad\qquad\qquad\qquad 
(T\not=T'\mbox{\ and\ } T\not=T'+1)
\end{array}
\end{displaymath}

Clearly, these three rules can be coded by a single Boolean circuit, and we have all the required Boolean logic at our disposal.

Because $\delta$ is a binary function, we need to compute a {\em cross product} on the coding of IDs to provide its input.
The transition function is therefore defined as:
\begin{displaymath}
\begin{array}{rcl}
\Phi &\equiv& \lambda p.\\
\ &\ &\quad\mbox{let }\langle u_1,u_2,u_3,u_4,u_5\rangle = \Copy_5\; p\mbox{ in}\\
\ &\ &\quad\mbox{let }\langle v_1,v_2,v_3.v_4,v_5\rangle = \Copy_5\; p\mbox{ in}\\
\ &\ &\qquad(\lambda w . w(\phi_T u_1 v_1) (\phi_S u_2 v_2) \dots (\phi_b u_5 v_5))\\
\ &\ &\qquad(\lambda w_T. \lambda w_S. \lambda w_H. \lambda w_C. \lambda w_b.\\
\ &\ &\quad\qquad w_T(\lambda z_1.\lambda z_2\dots\lambda z_T.\\
\ &\ &\qquad\qquad w_S(\lambda z_{T+1}.\lambda z_{T+2}\dots\lambda z_{T+S}.\\
\ &\ &\quad\qquad\qquad \dots\\
\ &\ &\qquad\qquad\qquad w_b(\lambda z_{C+1}.\lambda z_{C+2}\dots\lambda z_{C+b=m}.\\
\ &\ &\quad\qquad\qquad\qquad \lambda w.wz_1z_2\dots z_m)\dots)))
\end{array}
\end{displaymath}
The {\tt Copy} functions just copy enough of the input for the separate calculations to
be implemented in a linear way.
Observe that this $\lambda$-term is entirely linear {\em except} for
the two occurrences of its parameter $p$.  In that sense, it serves a
function analogous to $\lambda x.\Implies\, x\, x$ in the toy
calculation.  Just as $x$ ranges there over the closures for $\True$
and for $\False$, $p$ ranges over all possible IDs flowing to the
argument position.  Since there are two occurrences of $p$, we have
two entirely separate iterations in the $k$CFA analysis.  These
separate iterations, like nested ``for'' loops, create the equivalent
of a cross product of IDs in the ``inner loop'' of the flow analysis.

\subsection{Context and widget}

The context for the Turing machine simulation needs to set up the
initial ID and associated machinery, extract the Boolean value telling
whether the machine accepted its input, and feed it into the flow
widget that causes different flows depending on whether the value flowing in is
$\True$ or $\False$.  The following context is used for these purposes:
\begin{eqnarray*}
C &\equiv & (\lambda f_1.(f_1\;\Zero)(f_1\;\One))\\
\ &\ & (\lambda z_1.\\
\ &\ & \quad(\lambda f_2.(f_2\;\Zero)(f_2\;\One))\\
\ &\ & \quad(\lambda z_2.\\
\ &\ & \qquad\cdots\\
\ &\ & \quad\qquad(\lambda f_N. (f_N\;\Zero) (f_N\;\One))\\
\ &\ & \quad\qquad(\lambda z_N.((\lambda x.x) (\Widget (\Extract[\;]))^\ell)^{\ell'})\cdots)).
\end{eqnarray*}
In this code, the $\lambda x.x$ (with label $\ell'$ on its application) serve as padding, so that the term within is always applied in the same contour.
\Extract\ extracts a final ID, with its time stamp, and checks if it
codes an accepting state, returning \True\ or \False\ accordingly.
\Widget\ is our standard control flow test.  The context is
instantiated with the coding of the transition function, iterated over
an initial machine ID,
\begin{displaymath}
\begin{array}{l}
C[2^n\;\Phi\; \lambda w.w\Zero\dots\Zero\cdots Q_0\cdots H_0\cdots z_1 z_2\dots z_N \Zero],
\end{array}
\end{displaymath}
where $\Phi$ is a coding of transition function for $M$.  The
$\lambda$-term $2^n$ is a fixed point operator for $k$CFA, which can
be assumed to be either $\mathbf{Y}$, or an exponential function
composer.  There just has to be enough iteration of the transition
function to produce a fixed point for the flow analysis.

To make the coding easy, we just assume that $M$ starts by writing $x$ on the
tape, and then begins the generic exponential-time computation.  Then we can just have
all zeroes on the initial tape configuration.

\begin{lemma}
  For any Turing machine $M$ and input $x$ of length $n$, where $M$
  accepts or rejects $x$ in $2^{p(n)}$ steps, there exists a
  logspace-constructible, closed, labeled $\lambda$-term $e$ with distinguished label
  $\ell$ such that in the $k$CFA analysis of $e$ ($k>0$), $\True$ flows into
  $\ell$ iff $M$ accepts $x$.
\end{lemma}
\begin{theorem}
Deciding the control flow problem for $k$CFA with $k>0$ is complete
for \exptime.
\end{theorem}

\subsection{Exactness and PTIME}

At the heart of the \exptime-completeness result is the idea that the
{\em approximation} inherent in abstract interpretation is being
harnessed for computational power, quite apart from the power of {\em
exact} normalization.  To get a good lower bound, this is necessary:
it turns out there is a dearth of computation power when $k$CFA
corresponds with normalization, i.e.~when the analysis is exact.

As noted earlier, approximation arises from the truncation of contours
during analysis.  Consequently, if truncation never occurs, the
instrumented interpreter and the abstract interpreter produce
identical results for the given program.  But what can we say about
the complexity of these programs?  In other words, what kind of
computations can $k$CFA analyze exactly when $k$ is a constant,
independent of the program analyzed?

An answer to this question provides another point in the
characterization of the expressiveness of an analysis.  For 0CFA, the
answer is \ptime\ since the evaluation of linear terms is captured.
For $k$CFA, the answer remains the same: for any fixed $k$, $k$CFA can
only analyze polynomial time programs exactly.  It is only through the
use of approximation that a exponential time computation can be
simulated, but this computation has little to do with the actual
running of the program.  A program that runs for exponential time
cannot be analyzed exactly by any fixed $k$CFA.  Contrast this with
ML-typability, for example, where the evaluation of programs that run
for exponential time can be simulated via type inference.

Note that if the contour is never truncated, every program point is
now approximated by at most one closure (rather than an exponential
number of closures).  The size of the cache is then bounded by a
polynomial in $n$; since the cache is computed monotonically, the
analysis and the natural related decision problem is constrained by
the size and use of the cache.

\begin{proposition}
Deciding the control flow problem for exact $k$CFA is complete for
\ptime.
\end{proposition}

This proposition provides a characterization of the computational
complexity (or expressivity) of the language evaluated by the
instrumented evaluator $\mathcal{E}$ of section \ref{sec:instrumented}
as a function of the contour length.

It also provides an analytic understanding of the empirical
observation researchers have made: computing a more precise analysis
is often cheaper than performing a less precise one, which ``yields
coarser approximations, and thus induces more merging. More merging
leads to more propagation, which in turn leads to more reevaluation''
\cite{wright-jagannathan-toplas98}.  \citet{might-shivers-icfp06} make
a similar observation: ``imprecision reinforces itself during a flow
analysis through an ever-worsening feedback loop.''  This
ever-worsening feedback loop, in which we can make \False\
(spuriously) flow out of $\Implies\,x\,x$, is the critical ingredient
in our \exptime\ lower bound.

Finally, the asymptotic differential between the complexity of exact
and abstract interpretation shows that abstract interpretation is
strictly more expressive, for any fixed $k$.

\subsection{Discussion}

We observe an ``exponential jump'' between contour length and
complexity of the control flow decision problem for every
polynomial-length contour, including contours of constant length.
Once $k=n$ (contour length equals program size), an exponential-time
hardness result can be proved which is essentially a linear circuit
with an exponential iterator---very much like \cite{mairson-popl90}.
When the contours are exponential in program length, the decision
problem is doubly exponential, and so on.

The reason for this exponential jump is the cardinality of
environments in closures.  This, in fact, is the bottleneck for
control flow analysis---it is the reason that 0CFA (without closures)
is tractable, while 1CFA is not.  If $f(n)$ is the
contour length and $n$ is the program length, then
\begin{displaymath}
|\CEnv| = |\Var \rightarrow \Delta^{\leq f(n)}| = (n^{f(n)})^n = 2^{f(n)n \lg n}
\end{displaymath}
This cardinality of environments effectively determines the size of
the universe of values for the abstract interpretation realized by CFA.

When $k$ is a constant, one might ask why the inherent complexity is
exponential time, and not more---especially since one can iterate (in
an untyped world) with the $\mathbf{Y}$ combinator.  Exponential time
is the ``limit'' because with a polynomial-length tuple (as
constrained by a logspace reduction), you can only code an exponential
number of closures.

Finally, we need to emphasize the importance of linearity in static
analysis.  Static analysis makes approximations to be tractable, but
with linear terms, there is not approximation.  We carefully admitted
a certain, limited nonlinearity in order to increase the lower bound.

\section{Related work}

Our earlier work on the complexity of compile-time type inference is a
precursor of the research insights described here, and naturally so,
since type inference is a kind of static analysis.  The decidability
of type inference depends on the making of approximations, necessarily
rejecting programs without type errors; in simply-typed
$\lambda$-calculus, for instance, all occurrences of a variable must
have the same type.  (The same is, in effect, also true for ML, modulo
the finite development implicit in {\tt let}-bindings.)  The type
constraints on these multiple occurrences are solved by first-order
unification.

As a consequence, we can understand the inherent complexity of type
inference by analyzing the expressive power of {\em linear} terms,
where no such constraints exist, since linear terms are always
simply-typable.  In these cases, type inference is synonymous with
normalization.\footnote{An aberrant case of this phenomenon is
analyzed in \citet{neergaard-mairson-icfp04}, which analyzed a type
system where normalization and type inference are synonymous in
{\em every} case.  The tractability of type inference thus implied a
certain inexpressiveness of the language.}  This observation motivates
the analysis described in \citet{mairson-popl90, mairson-jfp04}.

The intuition behind the correspondence between evaluation and flow
analysis for linear terms can be seen as an instance of {\em abstract
counting} in the extreme \cite{might-shivers-icfp06}.  Abstract
counting is a technique for reasoning about the behavior of a program
that {\em must} occur when a program is run, based solely on abstract
information that describes what {\em may} occur.  When an
abstract value is a singleton set, the abstract object is effectively
rendered concrete \cite{jagannathan-etal-popl98}.  In other words,
when only one thing may happen, it must.  Linearity maintains
singularity, and analysis is therefore completely concrete.

Our coding of Turing machines is descended from earlier work on
Datalog (Prolog with variables, but without constants or function
symbols), a programming language that was of considerable interest to
researchers in database theory during the 1980s; see
\citet{hkmv,gmsv}.

In $k$CFA and abstract interpretation more generally, an expression
can evaluate to a set of values from a finite universe, clearly motivating
the idiom of programming with sets.  Relational database queries
take as input a finite set of tuples, and compute new tuples from
them; since the universe of tuples is finite and the computation is
monotone, a fixed-point is reached in a finite number of iterations.  The
machine simulation here follows that framework very closely.  Even the
idea of splitting a machine configuration among many tuples has its
ancestor in \cite{hkmv}, where a ternary ${\tt cons}(A,L,R)$ is used
to simulate a {\tt cons}-cell at memory address $A$, with pointers
$L,R$.  It needs emphasis that the computing with sets described in
this paper has little to do with normalization, and everything to do
with the approximation inherent in the abstract interpretation.

This coding of Boolean logic in linear $\lambda$-calculus, which was
previously given in \citet{vanhorn-mairson-icfp07} and is briefly described
again here for completeness, improves upon
\citet{mairson-jfp04} in that it allows uniform typing, and does not create garbage.
The encoding in \citet{mairson-jfp04} in turn is an improvement of the
Church encodings in that they are linear and non-affine.

Although $k$CFA and ML type inference are two static analyses complete
for \exptime\ \cite{mairson-popl90}, the proofs of these respective
theorems is fundamentally different.  The ML proof relies on type
inference simulating exact normalization (analogous to the
\ptime-completeness proof for 0CFA), hence subverting the approximation
of the analysis.  In contrast, the $k$CFA proof harnesses the
approximation that results from nonlinearity.

Recent work by \citet{might-phd} and \citet{might-shivers-icfp06} has
examined various techniques for reducing the imprecision of flow
analysis via {\em abstract garbage collection} and {\em abstract
counting}.  Might and Shivers observe that by eliminating spurious
flows, not only is the precision improved, but this often leads to
improved running times of the analyzer.  Our theorems reinforce these
observations and shed light on what might otherwise seems like a
paradoxical situation: ``in many cases, higher speed is a direct
consequence of higher precision.''  When the analysis is at its most
precise, it is always computable quickly.  On the other hand, wielding
the full power of spurious flows results in \exptime-completeness.  In
essence, $k$CFA is hard {\em because of} the spurious flows that it
must compute.  In this light, techniques such as abstract garbage
collection undermine our lower bound proofs, making it unclear what
complexity bounds exist for these enhanced analyses.

\section{Conclusions and perspective}

Empirically observed increases in costs can be understood analytically
as {\em inherent in the approximation problem being solved}.

We have given an exact characterization of the $k$CFA approximation
problem.  The \exptime\ lower bound validates empirical observations
and proves there is no tractable algorithm for $k$CFA.

The proof relies on previous insights about linearity, static
analysis, and normalization (namely, when a term is linear, static
analysis and normalization are synonymous); coupled with new insights
about using non-linearity to realize the full computational power of
approximate, or abstract, interpretation.

Shivers wrote in his best of PLDI retrospective
\citeyearpar{shivers-sigplan04},
\begin{quote}\em
Despite all this work on formalising CFA and speeding it up, I have
been disappointed in the dearth of work extending its {\em
power}.
\end{quote}
This work has shown that work spent on speeding up $k$CFA is an
exercise in futility; there is no getting around the exponential
bottleneck of $k$CFA.  The one-word description of the bottleneck is {\em closures},
which do not exist in 0CFA, because free variables in a closure would necessarily map to $\epsilon$,
and hence the environments are useless.

As for extending its power, from a complexity
perspective, we can see that 0CFA is strictly less expressive than
$k$CFA.  In turn, $k$CFA is strictly less expressive than, for example, Mossin's
flow analysis \citeyearpar{mossin-sas97}.  Mossin's analysis is a
stronger analysis in the sense that it is exact for a larger class of
programs than 0CFA or $k$CFA---it exact not only for linear terms, but
for all simply-typed terms.  In other words, the flow analysis of
simply-typed programs is synonymous with running the program, and
hence non-elementary.  This kind of expressivity is also found in
Burn-Hankin-Abramsky style strictness analysis
\citeyearpar{burn-hankin-abramsky}.  But there is a considerable gap
between $k$CFA and these more expressive analyses.  What is in between
and how can we build a real {\em hierarchy} of static analyses that
occupy positions within this gap?

L\'evy's notion of labeled reduction \citeyearpar{levy-phd} provides a
richer notion of ``instrumented evaluation'' coupled with a richer
theory of exact flow analysis, namely the geometry of interaction.
With the proper notion of abstraction and simulated reduction, we
should be able to design more powerful flow analyses, filling out the
hierarchy from 0CFA up to the expressivity of Mossin's analysis in the
limit.

\paragraph{Acknowledgments}
Thanks to Matt Might, Olin Shivers, and Mitch Wand for listening to
preliminary presentations of this proof.

\bibliographystyle{plainnat}


\end{document}